\documentclass{aa}
\usepackage{txfonts}
\usepackage{graphicx}
\usepackage{caption}
\usepackage{subcaption}

\usepackage{color}

\begin{document} 

\titlerunning{Connection between AGN and SFR}
\authorrunning{Masoura et al.}

\title{Disentangling the AGN and star formation connection using XMM-Newton}

\author{V. A. Masoura\inst{1,2}, G. Mountrichas\inst{1}, I. Georgantopoulos\inst{1}, A. Ruiz\inst{1}, G. Magdis\inst{3}, M. Plionis\inst{1,2}}
          
    \institute {National Observatory of Athens, V.  Paulou  \& I.  Metaxa, 11532,  Greece
              \email{vmasoura@noa.gr}
           \and
             Section of Astrophysics, Astronomy and Mechanics, Department of Physics, Aristotle University of Thessaloniki, 54 124, Thessaloniki, Greece
           \and 
             Dark Cosmology Centre, Niels Bohr Institute, University of Copenhagen, Juliane Mariesvej 30, DK-2100 Copenhagen, Denmark}

\abstract {There is growing evidence supporting the coeval growth of galaxies and their resident super-massive black hole (SMBH). Most studies also claim a correlation between the activity of the SMBH and the star formation of the host galaxy. It is unclear, however, whether this correlation extends to all redshifts and X-ray luminosities. Some studies find a weaker dependence at lower luminosities and/or a suppression of the star formation at high luminosities. We here use data from the X-ATLAS and XMM-XXL North fields and compile the largest X-ray sample up to date to investigate how X-ray selected AGN affect the star formation of their host galaxies in a wide redshift and luminosity baseline of  $0.03<z<3$ and $\rm log~L_X (\rm 2-10~keV)= (41-45.5)\,erg~s^{-1}$. Our sample consists of 3336 AGN. 1872 of our sources have spectroscopic redshifts. For the remaining sources we calculate photometric redshifts using TPZ, a machine-learning algorithm. We estimate stellar masses (M$_*$) and star formation rates (SFRs) by applying spectral energy distribution fitting through the CIGALE code, using optical, near-IR, and mid-IR photometry (SDSS, VISTA, and WISE). Of our sources, 608 also have far-IR photometry (Herschel). We use these sources to calibrate the SFR calculations of our remaining X-ray sample. Our results show a correlation between the X-ray luminosity (L$_X$) and the SFR of the host galaxy at all redshifts and luminosities spanned by our sample. We also find a dependence of the specific SFR (sSFR) on redshift, while there are indications that the X-ray luminosity enhances the sSFR even at low redshifts. We then disentangle the effects of stellar mass and redshift on the SFR and again study its dependence on the X-ray luminosity. Towards this end, we estimate the SFR of main-sequence galaxies that have the same stellar mass and redshift as our X-ray AGN and compare them with the SFR of our X-ray AGN. Our analysis reveals that the AGN enhances the star formation of its host galaxy when the galaxy lies below the main sequence and quenches the star formation of the galaxy it lives in when the host lies above the main sequence. Therefore, the effect of AGN on the SFR of the host galaxy depends on the location of the galaxy relative to the main sequence.}


\maketitle

\section{Introduction}

~~~~Most, if not all, galaxies host a super-massive black hole (SMBH) in their centre. The mass of this SMBH is correlated with the properties of its bulge, parametrised by the luminosity \citep{Magorrian2008} or the velocity dispersion \citep{Ferrarese2000}. When the material surrounding the galaxy is enough for the black hole to be fed, the galaxy is called active and its centre is called  an active galactic nucleus (AGN). The AGN are among the most luminous persistent sources in the Universe. Over the past 20 years, evidence  has been growing that supports the hypothesis of coeval growth of the galaxies and their resident SMBH \citep{Boyle1998}. This suggests some causal connection between the AGN and the star formation properties of the host galaxy \citep{Alexander2012}. 

Theoretical and semi-analytical models of galaxy evolution, through mergers, assume such a connection, where AGN feedback (the process by which the SMBH may moderate the growth of its host) plays a catalytic role \citep{Hopkins2006a, DiMatteo2008}. A model that has received particular interest proposes that AGN are triggered by mergers \citep{Hopkins2008a}. According to this model, the main phase of AGN growth coincides with host activity, which is likely to obscure the AGN. However, eventually, the powerful AGN pushes away the surrounding star-forming  material, arresting further star formation (quenching) and revealing the now-unobscured AGN. Different physical processes have been proposed to provide this quenching. These mechanisms can be broadly divided into two categories \citep{Gabor2010}. Those that heat gas that then cannot collapse to form stars (preventative feedback), for instance, virial shock heating \citep[e.g.][]{Birnboim2003}, galaxy interactions \citep[e.g.][]{DiMatteo2005}, and those that expel the gas that could be used to form stars \citep[ejective feedback,][]{Keres2009}. Different mechanisms prevail at high and low redshifts \citep{Hopkins2010}. Therefore observational constraints on the AGN luminosity in relation to the star formation activity can place important constraints on the theoretical models. 

The most straightforward way to study the effect of AGN on the star-formation of its host galaxy is to measure the star formation rate (SFR) as a function of the X-ray luminosity, L$_X$. X-rays detect the activity of the central SMBH, and therefore the X-ray luminosity is often used as a proxy of the AGN power \citep[e.g.][]{Lusso2012}. Observations at X-ray wavelengths provide a quite efficient way of selecting AGN over a wide luminosity baseline nearly independently of obscuration. Infrared observations provide a nearly uncontaminated view of star formation, as these long wavelengths are dominated by dust emission from the host galaxy. 

Several works have been described the correlation between the AGN X-ray luminosity and the SFR. Some groups have reported a strong link between star formation and AGN activity for high-luminosity AGN, but at lower luminosities, these correlations appear relatively weak or absent \citep[e.g.][]{Lutz2010, Bonfield2011}. \cite{Page2012} used data from CDFN and found a positive correlation of SFR with L$_X$ for $\rm log~L_X (\rm 2-10~keV)<44 \,erg~s^{-1}$.  However, their analysis reveals a suppression of star formation at higher X-ray luminosities. This quenching is in agreement with theoretical models in which the AGN outflows expel the interstellar medium of the host galaxy \citep{DiMatteo2005, Springel2005, Sijacki2007}. This possible suppression of the star formation due to the AGN activity may also be the reason for the galaxies' transition from the blue to the red cloud \citep{Georgakakis2008}. \cite{Harrison2012} combined sources from CDFN, CDFS, and the COSMOS fields and revealed a dependence of the star formation on redshift. However, they did not detect any statistical significant correlation between SFR and X-ray luminosity. Most of these studies suffer from low number statistics, however \citep[see][]{Harrison2012}. \cite{Brown2018} used 703 AGN with $\rm log~L_X (\rm 2-10~keV)= (42-46)\,erg~s^{-1}$ at 0.1 < z < 5 from the Chandra XBo$\ddot{\rm{o}}$tes X-ray Survey and found a dependence of the star formation of the host galaxy on X-ray luminosity. \cite{Lanzuisi2017} used 692 AGN from the COSMOS field in the redshift range 0.1<z<4. They did detect a dependence of the SFR on X-ray luminosity at all redshifts and luminosities, but their SFR calculations appear discrepant compared to those from \cite{Brown2018}. 

The AGN-SFR relation can also be investigated by calculating the specific star formation rate (sSFR; defined as the ratio of SFR to M$_*$) and compare it with the L$_X$. \cite{Rovilos2012} used X-ray data from the 3Ms CDFS XMM-{\it{Newton}} survey and found a significant correlation between the sSFR and X-ray luminosity for redshifts $\rm{z}>1$. They did not detect a strong correlation at lower redshifts, however.

We address the contradictory results from previous studies by compiling multiwavelength data from both the XMM-XXL and X-ATLAS surveys. Our goal is to study the effect of AGN on the SFR of the host galaxy using the largest X-ray sample up to date. Towards this end, in Sections 4.1 and 4.2 we measure the SFR and sSFR as a function of X-ray luminosity, respectively. In Section 4.3 we examine the evolution of SFR with stellar mass and redshift. Finally, in Section 4.4 we disentangle the effects of stellar mass and redshift on the SFR and study its dependence on X-ray luminosity. In the calculations that follow, we adopt cosmological parameters H$_0 = 70$\,km\,s$^{-1}$\,Mpc$^{-1}$, $\Omega_ M = 0.3, \Omega _\Lambda = 0.7$.

\begin{figure*}
\centering
\begin{subfigure}{.52\textwidth}
  \centering
  \includegraphics[width=.9\linewidth]{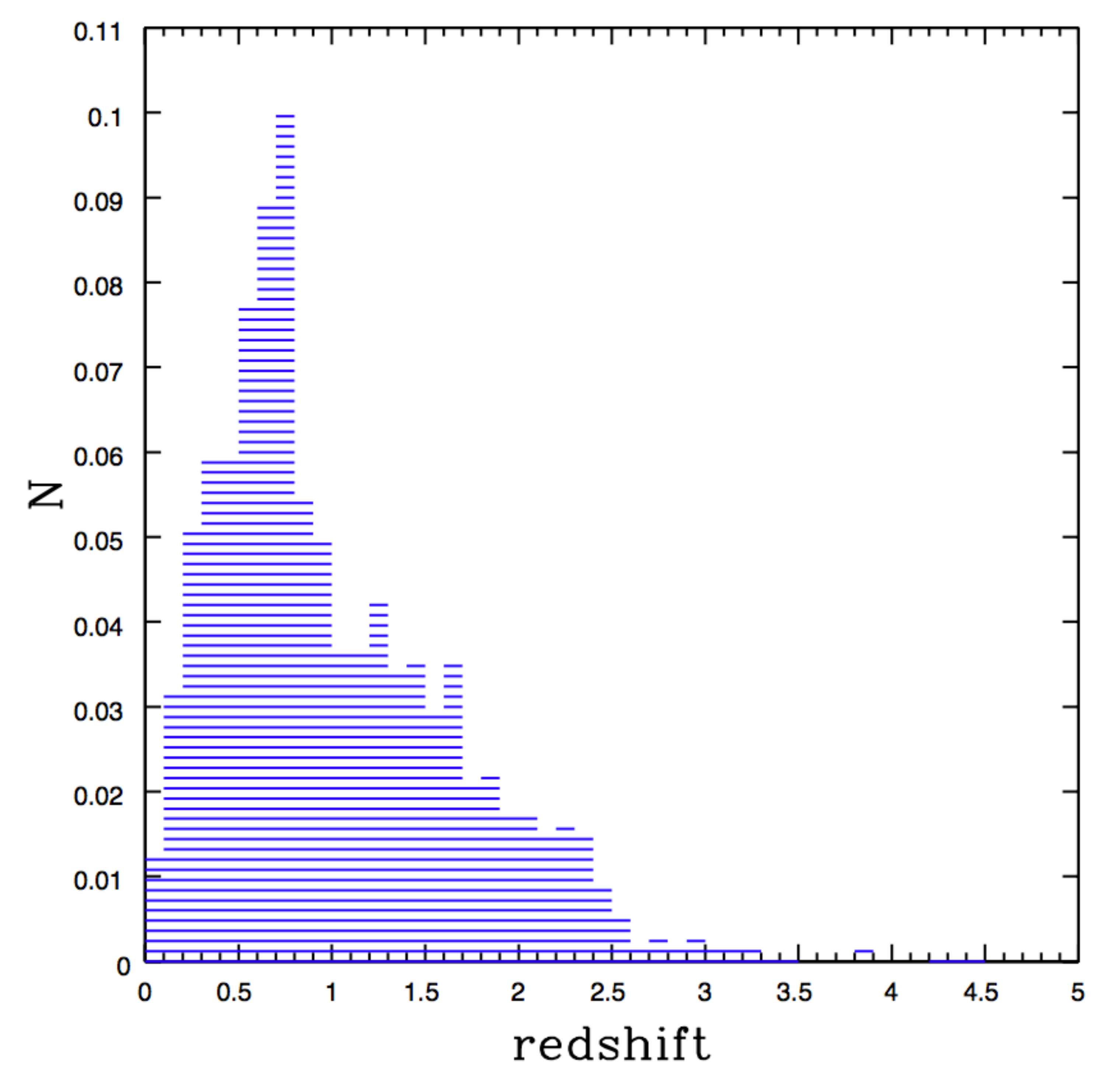}
  \label{fig:redz}
\end{subfigure}%
\begin{subfigure}{.52\textwidth}
  \centering
  \includegraphics[width=.9\linewidth]{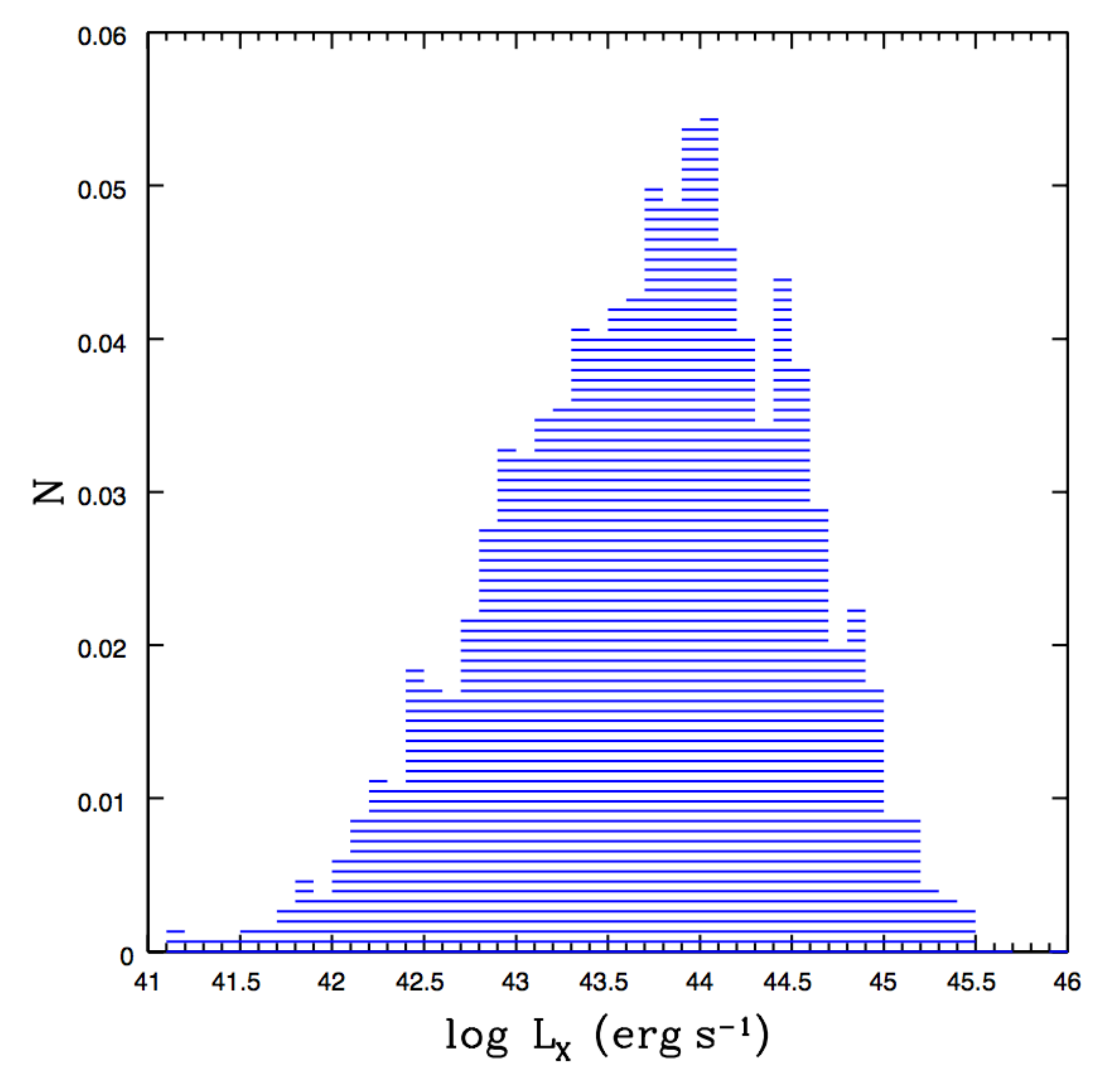}
  \label{fig:lx}
\end{subfigure}
\caption{Left: Redshift distribution of our 3336 X-ray sources. Right: X-ray luminosity distribution of the 3336 AGN. Both histograms have been normalised to the total number of sources. }
\label{redz_lx}
\end{figure*}

\begin{figure}
\includegraphics[height=1.\columnwidth]{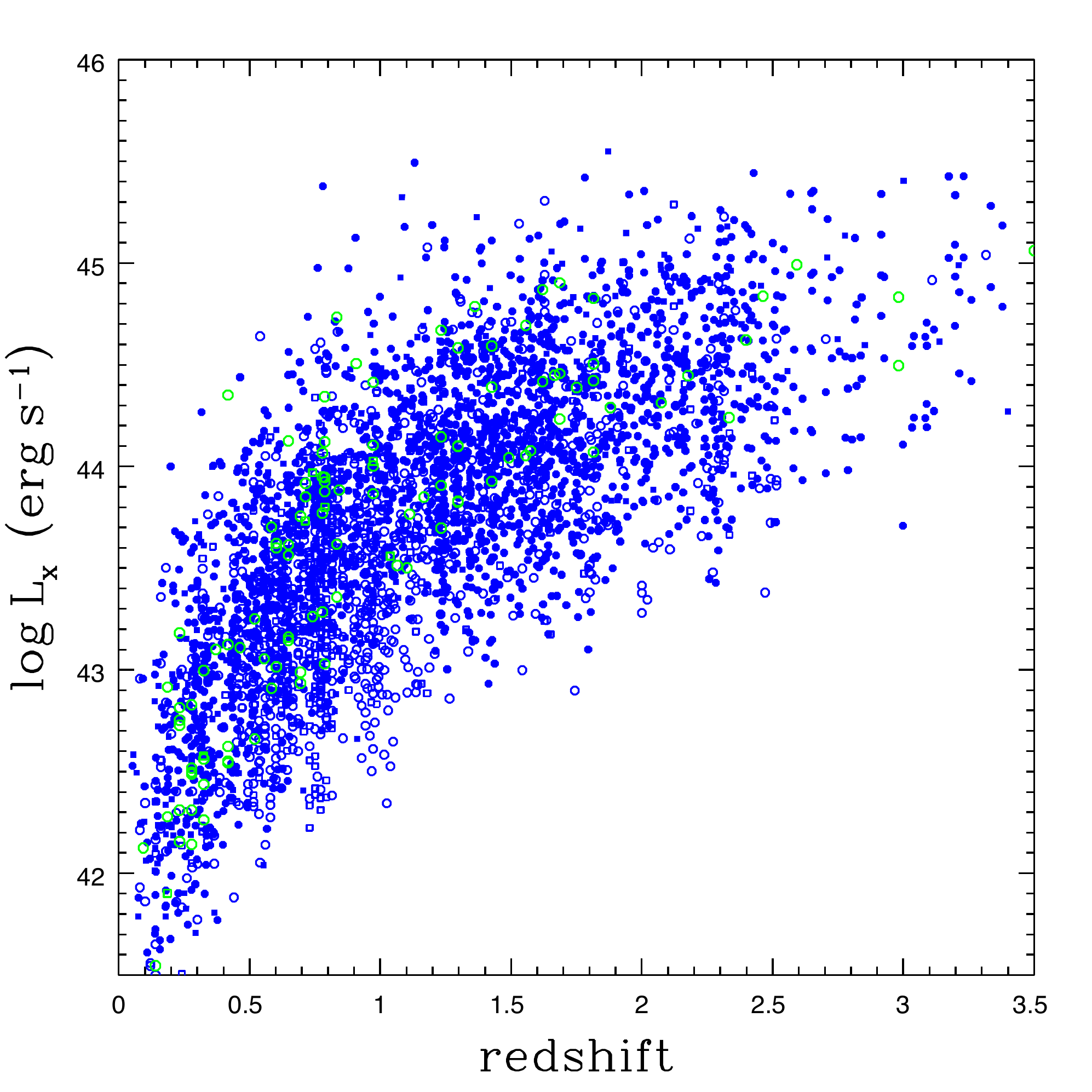}
\caption{The X-ray luminosity as a function of redshift for the 3336 X-ray AGN in our final sample. Sources in the XMM-XXL field are shown in blue. AGN in the X-ATLAS field are presented with green symbols. Sources with photoz are shown by empty points and specz sources by filled points. AGN with available Herschel photometry are presented with squares whereas those without Herschel by circles.}
\label{fig_lx_z}
\end{figure}

\begin{figure*}
\centering
\begin{subfigure}{.5\textwidth}
  \centering
  \includegraphics[width=1.\linewidth]{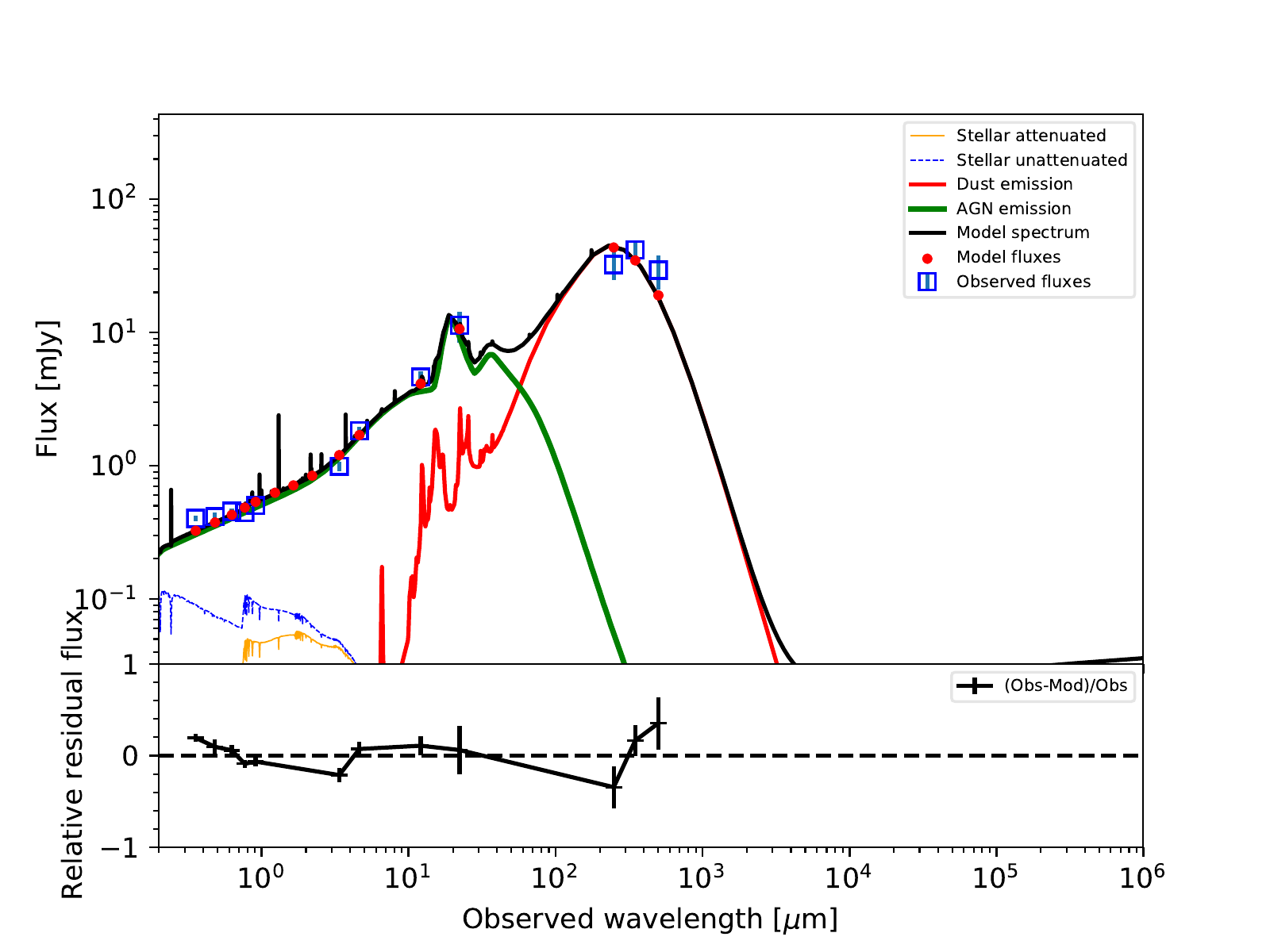}
  \label{fig:sub1}
\end{subfigure}%
\begin{subfigure}{.5\textwidth}
  \centering
  \includegraphics[width=1.03\linewidth]{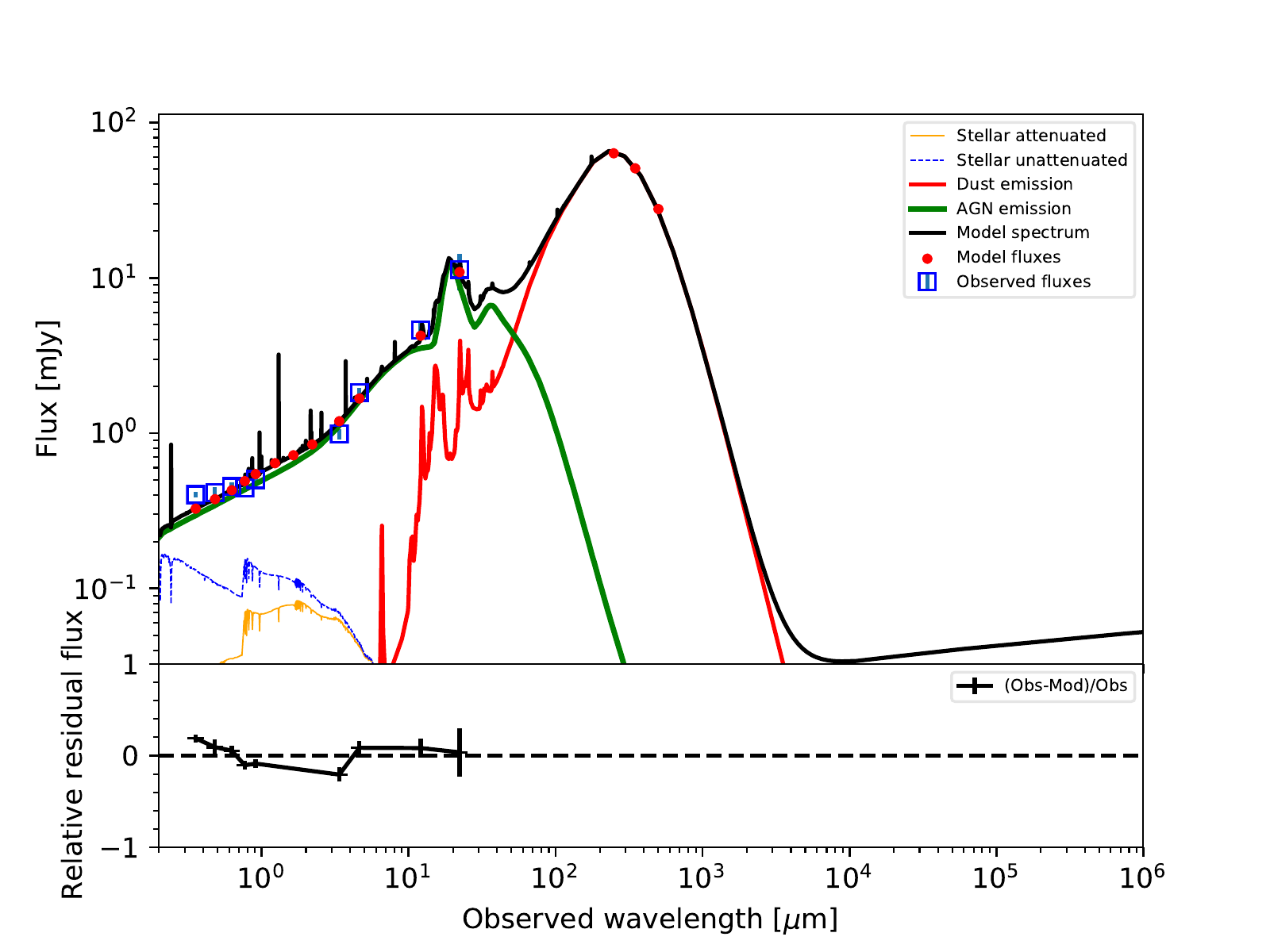}
  \label{fig:sub2}
\end{subfigure}
\caption{SED fitting of an AGN with (left panel) and without (right panel) HERSCHEL photometry. The star formation component is plotted in red, the AGN component in green, and the attenuated and the unattenuated stellar component is described by the yellow and the blue dashed line, respectively. The black solid line shows the best fit from CIGALE. The source lies at $\rm{z}=0.986$. When far-IR photometry is included in the fitting analysis, CIGALE yields $\chi _{red}^2=2.54$, log\,$\rm{SFR} =2.49\,M_\sun\,yr^{-1}$ and M$_\star=10.94\,M_\sun$. Without HERSCHEL, the corresponding values are $\chi _{red}^2=1.46$, log\,$\rm{SFR} =2.28\,M_\sun\,yr^{-1}$ , and M$_\star=11.05\,M_\sun$.}
\label{SED}
\end{figure*}

\section{Data}

In our analysis we use X-ray sources from two different fields: the XMM-XXL  and the X-ATLAS. In this section we describe how we compiled sources from the two datasets and the available photometry used to perform the SED fitting analysis. 

\subsection{X-ATLAS}

X-ATLAS is one of the largest contiguous areas of the sky with both XMM-Newton and Herschel coverage. The catalogue consists of 1816 X-ray AGN \citep{Ranalli2015}.  To obtain optical, mid-IR, and far-infrared photometry for these sources, we cross-matched the X-ray catalogue with the SDSS-DR13 \citep[u, g, r, i, z;][]{Albareti2015}, the WISE \citep[W1, W2, W3, W4;][]{Wright2010}, and the VISTA-VIKING \citep[J, H, K;][]{Emerson2006, Dalton2006} catalogues. The cross-match was performed using the ARCHES cross-correlation tool xmatch \citep{Pineau2017}. This tool symmetrically  matches an  arbitrary  number  of  catalogues  providing  a Bayesian probability of association or non-association. The cross-match revealed 1,031 sources with at least optical photometry \citep[for more details see][]{Mountrichas2017b}. 

To improve the accuracy of our SFR estimations, we also included far-IR photometry in our SED fitting analysis when available. For that purpose, we cross-matched the 1,031 sources with the Herschel Terahertz Large Area  sample (H-ATLAS).The H-ATLAS survey is the largest Open Time Key Project carried out with the Herschel Space Observatory \citep{Eales2010}, covering an area of 600 $deg^2$ in five far-infrared and sub-millimeter bands, 100, 160, 250, 350, and 500 $\mu$m \citep{Valiante2016}. Sixteen $deg^2$ have been presented in the Science Demonstration Phase (SDP) catalogue \citep{Rigby2011} and lie within one of the regions observed by the Galaxy And Mass Assembly (GAMA) survey \citep{Driver2011, Baldry2010}. Sixty-five sources are common between our dataset and the H-ATLAS catalogue.

\subsection{XMM-XXL}

The XMM-Newton XXL survey \citep[XMM-XXL;][]{Pierre2016} is a medium-depth X-ray survey that covers a total area of 50\,deg$^2$ split into two fields equal in size, the XMM-XXL North (XXL-N) and the XMM-XXL South (XXL-S). In our analysis, we used the XXL-N sample, which consists of 8445 X-ray sources. Of these X-ray sources, 5294 have SDSS counterparts and 2512 have reliable spectroscopy \citep{Menzel2016, Liu2016}. Mid-IR and near-IR was obtained following the likelihood ratio method \citep{Sutherland_and_Saunders1992} as implemented in \cite{Georgakakis_Nandra2011}. For more details on the reduction of the {\it{XMM}} observations and the IR identifications of the X-ray sources, see \cite{Georgakakis2017}.

The XXL field was partially observed by Herschel ($\sim$70\% of the XXL
area) in the context of the Herschel Multi-tiered Extragalactic Survey
\citep[HerMES,][]{Oliver2012}. We used the SPIRE xID250 catalogue from
the HERMES-DR4 (http://hedam.lam.fr/HerMES/index/dr4), a band-merged
catalogue (250, 350 and 500 $\mu$m) extracted on blind 250 $\mu$m positions.
We cross-matched this catalogue with the list of XXL X-ray sources
using xmatch.

To properly perform xmatch, all the crosmatched catalogues must
cover the same footprint. Before the crossmatch, we therefore selected
only X-ray and Herschel sources in the footprint resulting of the
intersection of the XXL and HERMES areas. There are 6790 X-ray sources
and 54\,823 Herschel sources in the common area. We used an average
positional error for the Herschel sources of 15 arcsec. After the
cross-match with xmatch, we rejected sources with a low probability of
association (<68\%). When the same X-ray source was
associated with several Herschel counterparts, we selected the
association with the highest probability. After applying these
filtering criteria, we found 608 X-ray sources with a Herschel
counterpart.

\subsection{Final sample}
\label{sec_final_sample}

We used only those X-ray sources that had the most accurate SFR estimates. For this purpose, we included only AGN with at least WISE or HERSCHEL photometry, in addition to optical photometry. We also excluded sources that have been optically classified as extended and their photometric redshift (see Section 3.1), photoz, is  photoz$>1$ \citep{Salvato2011}. Moreover, we excluded AGN for which their SED fitting has estimated a reduced $\chi ^2$, $\chi^2_{\rm{red}}>5$ (see Section 3.2). This criterion is based on visual inspection of the SED fits. The final number of AGN are presented in Table \ref{sample}. All our X-ray sources have  $\log \, L_X (\rm 2-10\,keV)  > 41.0\, erg~s^{-1}$ , which minimises contamination from inactive galaxies. The distributions of redshift and  X-ray luminosity are presented in Figures \ref{fig:redz} and \ref{fig:lx}, respectively, for our full X-ray sample.

Various selection biases are introduced in our X-ray sample: a requirement for optical and mid-IR photometry, use of photometric redshifts and data from wide area surveys compared to deeper fields (e.g. COSMOS, Chandra Deep Fields). It is not straightforward to define how these affect our measurements. For instance, the SDSS requirement biases our sample against low-luminosity sources, whereas the XXM-XXL allows us to include more high-luminosity AGN in our dataset. When presenting our measurements, we therefore also include estimates for the individual AGN, using different symbols and colours depending on available spectroscopy and/or photometry. Furthermore, we split our final sample into many redshift bins to minimise any effect of possible incompleteness in the Lx-redshift plane (e.g. Figures \ref{SFR}, \ref{sSFRLx}, and \ref{Mstellar}). These allow us to better assess whether and how selection biases affect our calculations. 

In Fig. \ref{fig_lx_z} we present the X-ray luminosity as a function of redshift for our final sample. Our dataset lacks low-luminosity sources ($\log \, L_X (\rm 2-10\,keV)  < 42.5\, erg~s^{-1}$) at $\rm{z}>1$. Restricting our X-ray catalogue to AGN with $\log \, L_X (\rm 2-10\,keV)  > 43.5\, erg~s^{-1}$ reduces the number of sources to 2067, but our sample is complete up to $\rm{z}\sim 2.5$. The effect of this incompleteness is studied in Section \ref{sec_main} and is shown in Fig. \ref{Main}, which presents the main results of this study.

\begin{table*}
\caption{Number of  AGN with spectroscopic (specz) and photometric (photoz) redshifts in the XMM-XXL and X-ATLAS fields. In parentheses we quote the number of X-ray sources with available Herschel photometry.}

\centering
\setlength{\tabcolsep}{2.3mm}
\begin{tabular}{cccc}
      \hline
& XMM-XXL & X-ATLAS&Total sources\\
       \hline
specz sources& 1849(338)&23(3)&1872(341)   \\
photoz sources&1364(262)&100(5)&1464(267) \\
\hline
Total sources&3213(600)&123(8)&3336(608)\\
\hline
\label{sample}
\end{tabular}
\end{table*}

\section{Analysis}

\subsection{Photometric redshifts}

To perform an SED fitting analysis, we need redshift information for our X-ray sources. As described in the previous section, 2,512/5,294 AGN in the XXM-XXL field have available spectroscopic redshifts (specz), while for the $\sim 1,000$ sources in the X-ATLAS field we use the photoz catalogue presented in \cite{Mountrichas2017b}. For the remaining 2,782 X-ray sources in the XMM-XXL field we estimated photometric redshifts using TPZ, a publicly available, machine-learning algorithm. The code and the technique it incorporates are described in detail in \cite{Kind2013}. In summary, the algorithm uses prediction trees and random forest techniques to generate photometric redshift probability density functions (PDFs). As an empirical technique it requires a dataset with reliable spectroscopy to train the algorithm before it is applied to our photometric X-ray sample. We used the same training sample that \cite{Mountrichas2017b} used to estimate photoz for the X-ray sources in the X-ATLAS field and followed their analysis (see their Section 3 for more details). Based on their results, the estimated photometric redshifts have a normalised absolute median deviation, $\rm{nmad}\approx0.06$, and a percentage of outliers, $\eta=10-14\%$, depending upon whether the sources are extended or point-like.

Although we estimate photometric redshifts for the total of the 2,782 AGN, in our SED fitting analysis we include only sources that have at least WISE or HERSCHEL photometry, in addition to optical photometry (see Section 2.3). Thus, sources whose photoz has been estimated using only optical photometry and therefore is less accurate \citep[see Table 3 in][]{Mountrichas2017b} were excluded from our measurements.

\subsection{SFR and stellar mass estimations}

In this section, we describe the analysis we followed to estimate the SFRs and the stellar mass of our X-ray sources. 

\subsubsection{SED fitting}

To calculate SFRs and stellar masses for our X-ray AGN samples, we used the CIGALE code version 0.12 \citep[Code Investigating GALaxy Emission;][]{Noll2009, Ciesla2015}. We provided CIGALE with multiwavelength photometry (see previous section) as well as a list of possible values for physical parameters related to the star formation history, dust attenuation, nebular emission, and AGN emission (Table \ref{table_cigale}). The \cite{Fritz2006} library of templates was used to model the AGN emission. The double exponentially decreasing (2$\tau$ dec) model was used to convolve star formation histories \citep{Ciesla2015}. Our SFR estimations assume that all the far-IR emission is due to dust., that is, the AGN contribution is ignored. However, this is not the case for powerful AGN \citep[e.g.][]{Symeonidis2016, Duras2017}. This introduces a maximum offset of 30\% to our SFR calculations \citep[for more details, see Section 3.2.3 in][]{Ciesla2015}. The stellar population synthesis was modelled using the \cite{Bruzual_Charlot2003} template, adopting the Salpeter template. \cite{Calzetti2000} and \cite{Dale2014} templates were used for the dust extinction law and the absorbed dust reemitted in the IR. Fig. \ref{SED} presents two examples of our SED fitting analysis.

\subsubsection{Using Herschel photometry to calibrate SFR estimates}

\begin{figure}
\includegraphics[height=1.\columnwidth]{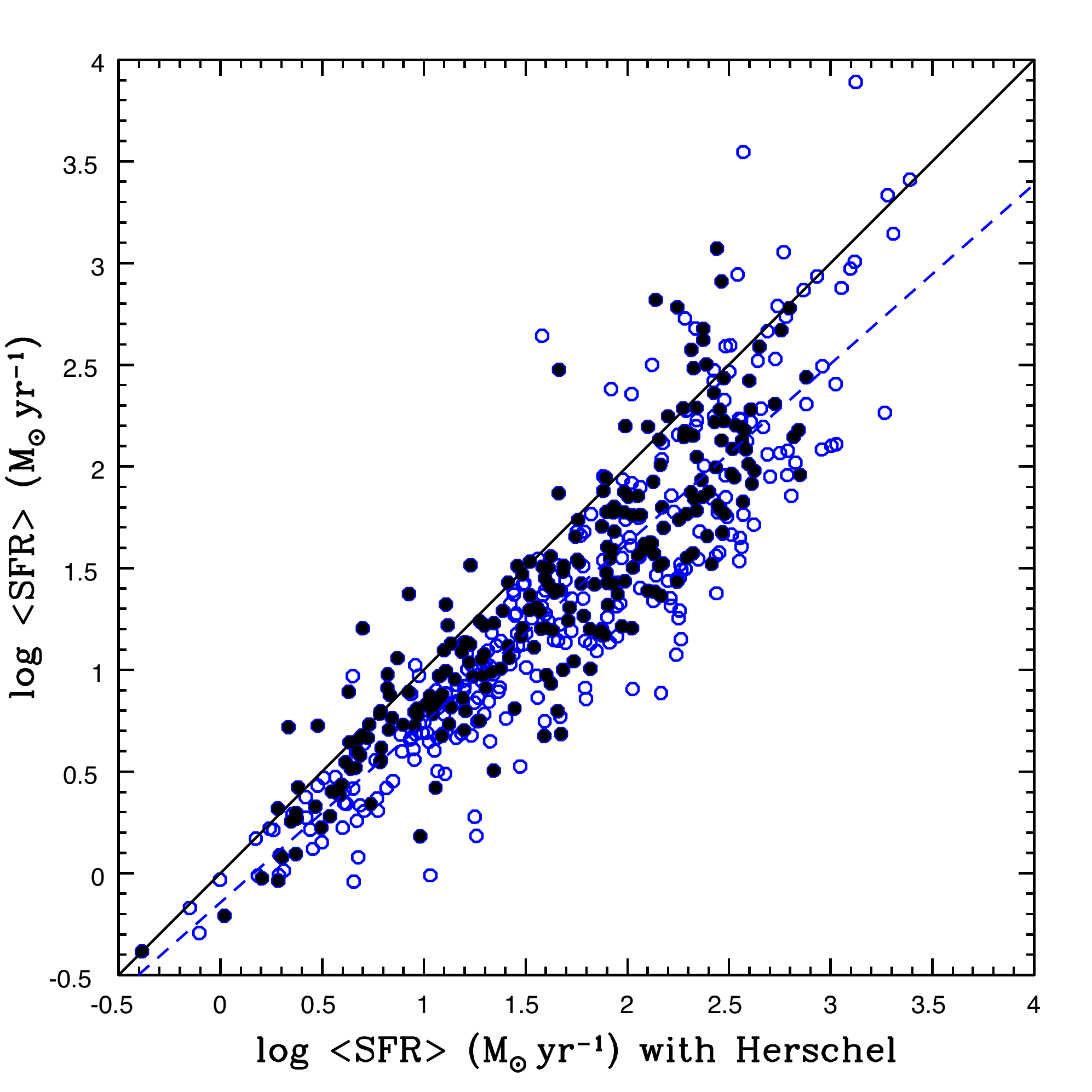}
\caption{Comparison of the SFR estimations with and without Herschel for the 608 sources with available Herschel photometry. The black solid line shows the 1-1 line, and the blue dashed line represents the best-fit calibration line. Sources with photoz are shown by blue, empty circles and specz sources by filled circles.}
\label{Fig_calibration}
\end{figure}

As shown in Table \ref{sample}, 608 out of the 3336 X-ray sources have additional Herschel photometry (PACS and SPIRE; 100, 160, 250, 350, and 500 $\mu$m). For these sources we performed an SED fitting following the analysis described in the previous section and estimated their SFR with and without the Herschel bands. In the latter case, only SDSS and WISE bands were used in the SED fitting. The comparison is shown in Fig. \ref{Fig_calibration}. Although there is a small scatter in the measurements that could be due to statistical errors (CIGALE estimations) and/or the usage of photometric redshifts for some sources, among others, we note a systematic offset in the measurements. SFRs estimated without Herschel are underestimated compared to SFR calculations including Herschel. Applying a $\chi^2$ fit, we find that this offset is best described by the following equation: 

\begin{equation}
\rm{log\,SFR_{Herschel}} = \frac{\rm{log\,SFR_{noHerschel}} + 0.12\pm0.01}{0.87\pm0.03}.
\end{equation}
The errors on the best-fit parameters represent the 1\,$\sigma$ uncertainties. Based on the above equation, the corresponding uncertainty on the calculated SFR$_{Herschel}$ due to the scatter is $<10\%$. Although in our analysis we use this equation to correct the SFR estimates for the sources that do not have far-IR photometry, we also separately present our results for sources with and without Herschel photometry. The inclusion of the (calibrated) non-Herschel SFR estimates significantly increases the size of our X-ray sample without negatively affecting our measurements or altering our conclusions.

\begin{figure*}
\centering
\begin{subfigure}{.5\textwidth}
  \centering
  \includegraphics[width=1.\linewidth]{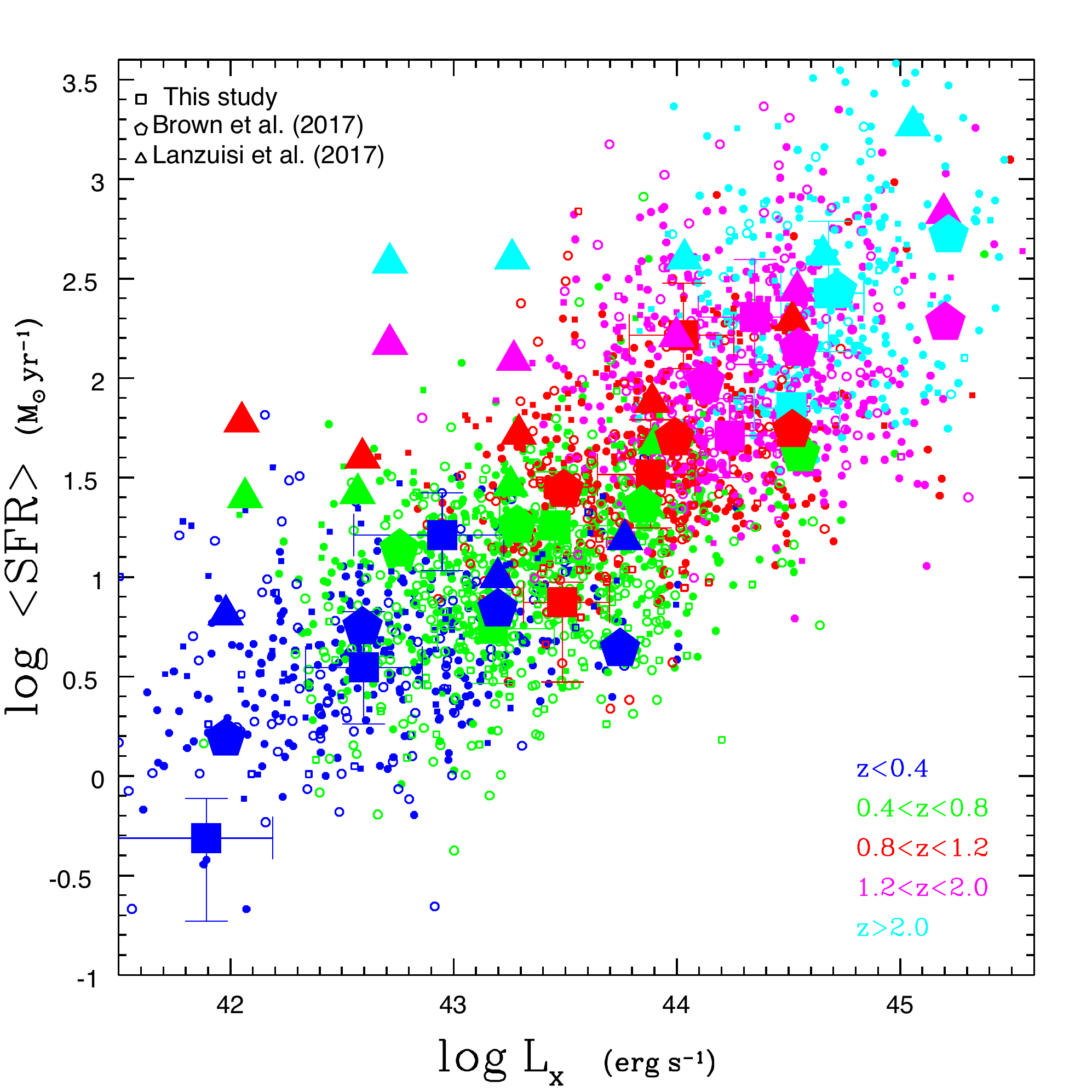}
  \label{fig:sub1}
\end{subfigure}%
\begin{subfigure}{.5\textwidth}
  \centering
  \includegraphics[width=1.03\linewidth]{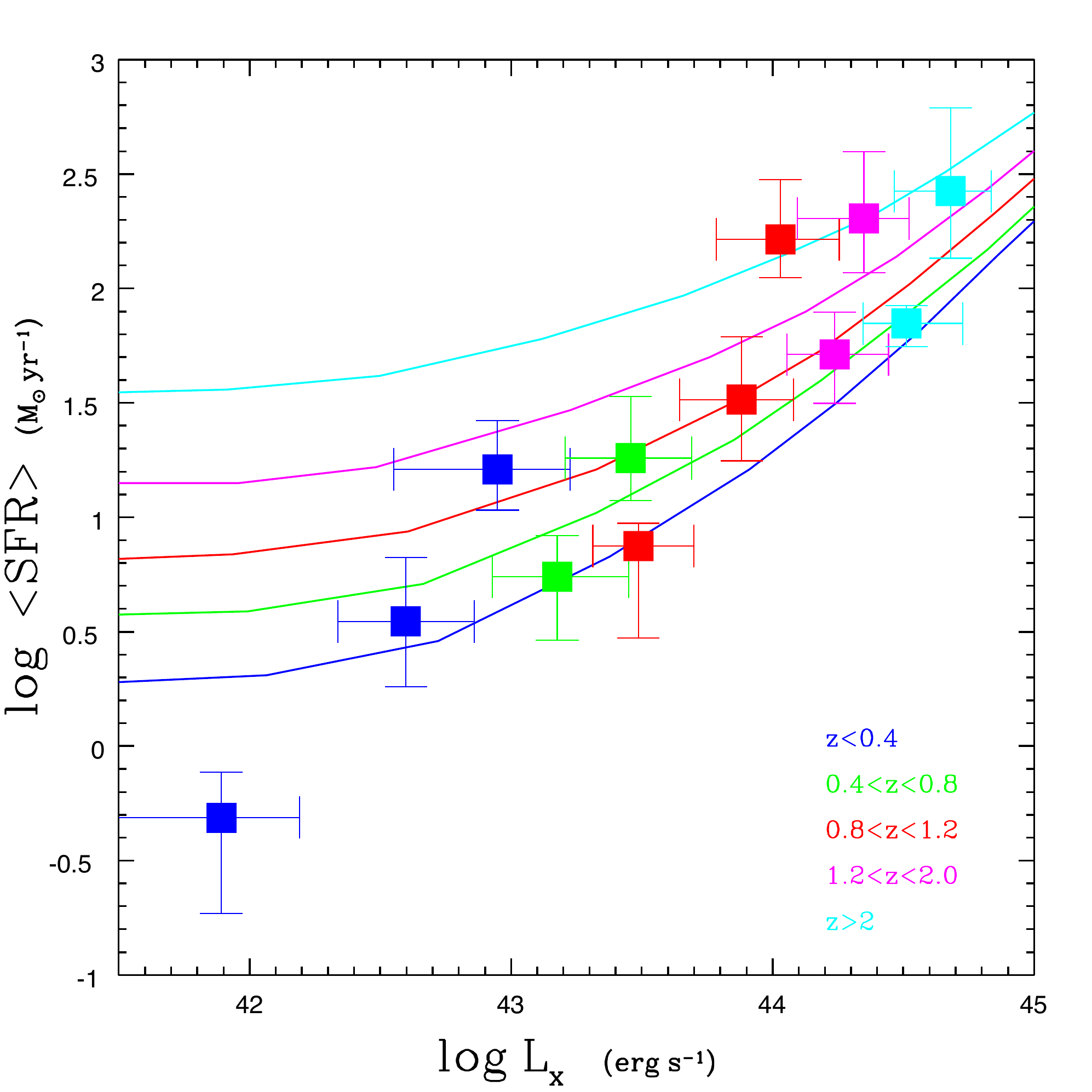}
  \label{fig:sub2}
\end{subfigure}
\caption{Left: Distribution of AGN X-ray luminosity vs. SFR. Dots show individual AGN. Sources with photoz are shown by empty points and specz sources by filled points. AGN with available Herschel photometry are presented with squares, and those without Herschel data by circles. Large squares refer to our binned results (median SFR and L$_X$ values are shown, in bins of SFR), and the error bars represent the 1$\sigma$ dispersion of each bin. Triangles and polygons show the results from \cite{Lanzuisi2017} and \cite{Brown2018}, respectively. The symbols are colour-coded based on their redshifts, z>0.4 blue, 0.4<z<0.8 green, 0.8<z<1.2 red, 1.2<z<2.0 magenta, and z>2.0 cyan. 
Right: Distribution of AGN X-ray luminosity vs. SFR. 
Squares present our binned results. The solid lines show the extrapolated trends from \cite{Hickox2014}. They are colour-coded based on their redshift range.}
\label{SFR}
\end{figure*}

\begin{table*}
\caption{Models and the values for their free parameters used by CIGALE for the SED fitting of our X-ray samples. $\tau$ is the e-folding time of the main stellar population model in Myr, age is the age of the main stellar population in the galaxy in Myr (the precision is 1\,Myr), and burst age is the age of the late burst in Myr (the precision is 1 Myr). $\beta$ and $\gamma$ are the parameters used to define the law for the spatial behaviour of density of the torus density. The functional form of the latter is $\rho (r, \theta) \propto  r^\beta e^{-\gamma | cos \theta |}$, where r and $\theta$ are the radial distance and the polar distance, respectively. $\Theta$ is the opening angle and $\Psi$ the viewing angle of the torus. Type-2 AGN have $\Psi=0$ and Type-1 AGN have $\Psi=90$. The AGN fraction is measured as the AGN emission relative to infrared luminosity ($1-1000\,\mu m$).} 
\centering
\setlength{\tabcolsep}{0.1mm}
\begin{tabular}{cc}
       \hline
Parameter &  Model/values \\
        \hline
\multicolumn{2}{c}{stellar population synthesis model} \\
\\
initial mass function & Salpeter \\
metallicity & 0.02 (Solar) \\
single stellar population library & Bruzual \& Charlot (2003) \\
\hline
\multicolumn{2}{c}{double exponentially decreasing (2$\tau$ dec) model} \\
$\tau$ & 100, 1000, 5000, 10000 \\ 
age & 500, 2000, 5000, 10000, 12000 \\
burst age & 100, 200, 400 \\
\hline
\multicolumn{2}{c}{Dust extinction} \\
dust attenuation law & \cite{Calzetti2000} \\
reddening E(B-V) & 0.01, 0.1, 0.3, 0.5, 0.8, 1.2 \\ 
E(B-V) reduction factor between old and young stellar population & 0.44 \\
\hline
\multicolumn{2}{c}{\cite{Fritz2006} model for AGN emission} \\ 
ratio between outer and inner dust torus radii &   10, 60, 150 \\
9.7 $\mu m$ equatorial optical depth & 0.1, 0.3, 1.0, 2.0, 6.0, 10.0 \\
$\beta$ & -0.5 \\
$\gamma$ & 0.0, 2.0, 6.0 \\
$\Theta$ & 100 \\
$\Psi$ & 0.001, 10.10, 20.10, 30.10, 40.10, 50.10, 60.10, 70.10, 80.10, 89.99 \\
AGN fraction & 0.1, 0.2, 0.3, 0.5, 0.6, 0.8 \\
\hline
\label{table_cigale}
\end{tabular}
\end{table*}

\section{Results and discussion}

\begin{figure}
\includegraphics[height=1.\columnwidth]{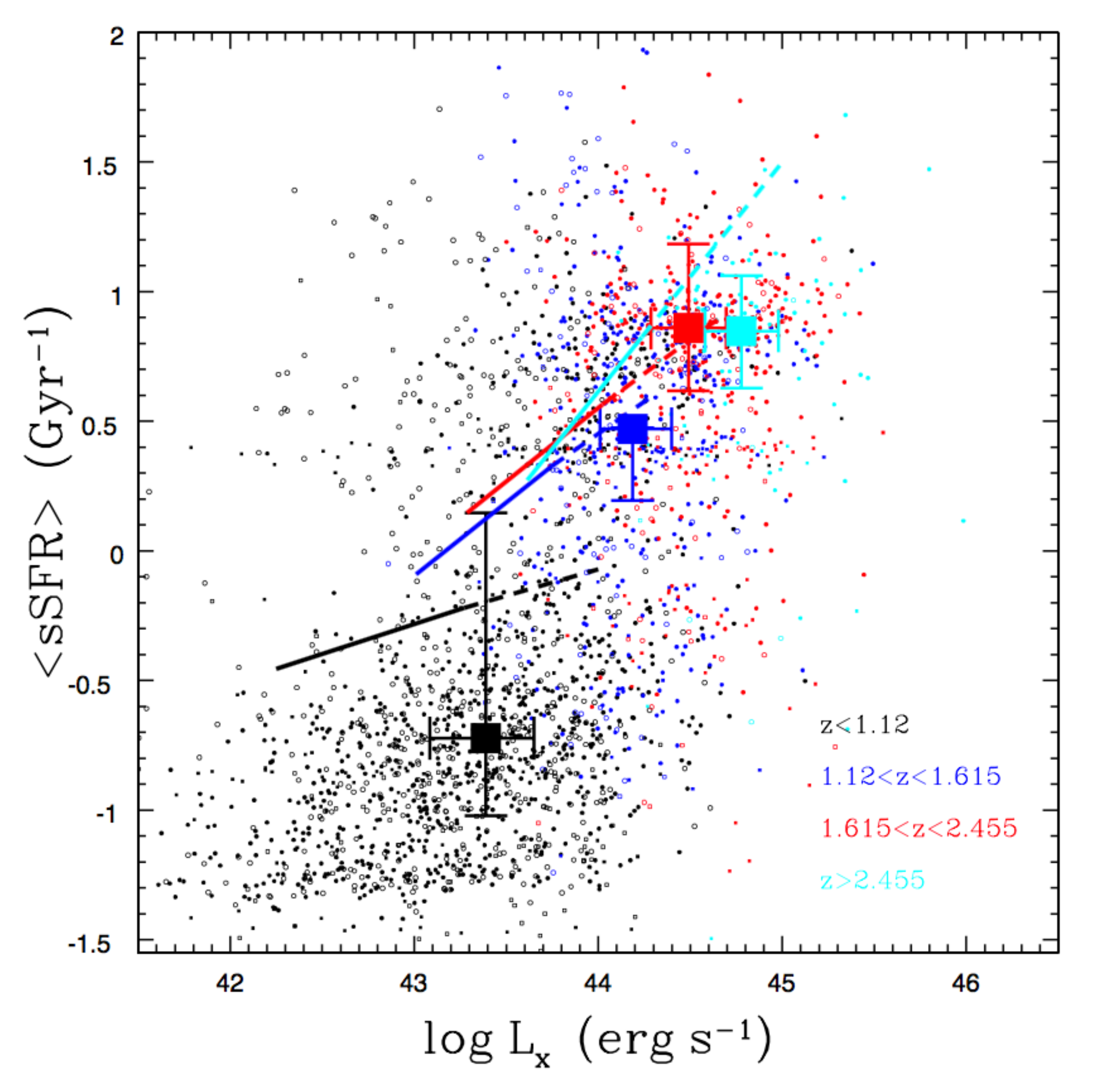}
\caption{Specific star formation rate against X-ray luminosity for our AGN sample. The black, blue, red, and cyan symbols refer to z < 1.120, 1.120 < z < 1.615, 1.615 < z < 2.455, and z > 2.455, respectively. Solid lines present the \cite{Rovilos2012} estimates, and the dashed lines show their extrapolation to higher luminosities. }
\label{sSFRLx}
\end{figure}

\begin{figure}
\includegraphics[height=.9\columnwidth]{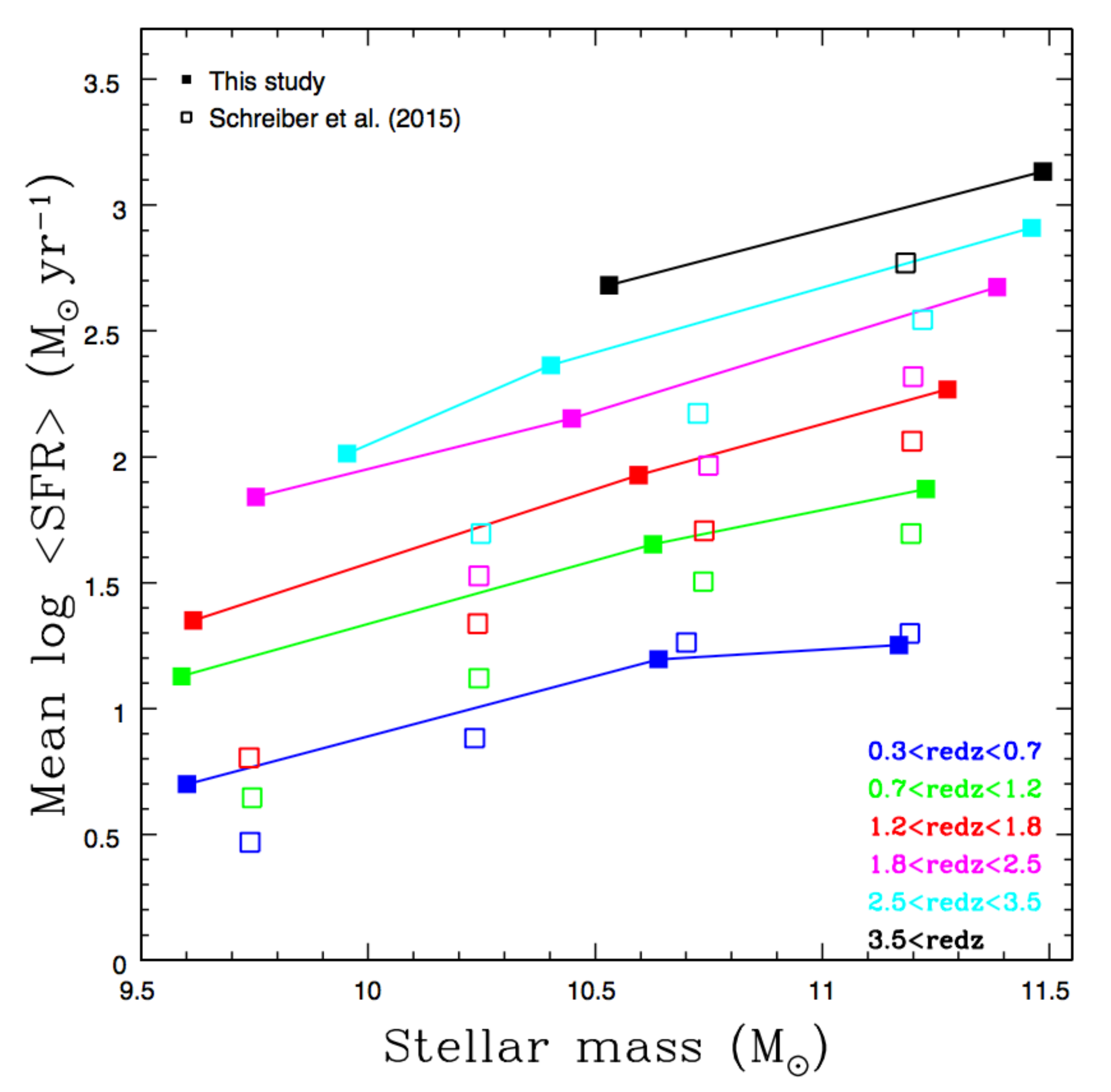}
\caption{Mean SFR versus $M_*$ for our sample (filled squares) and for observed star-forming galaxies \citep[open squares;][]{Schreiber2015}.}
\label{Mstellar}
\end{figure}


\begin{figure*}
\centering
\begin{subfigure}{.5\textwidth}
  \centering
  \includegraphics[width=1.\linewidth]{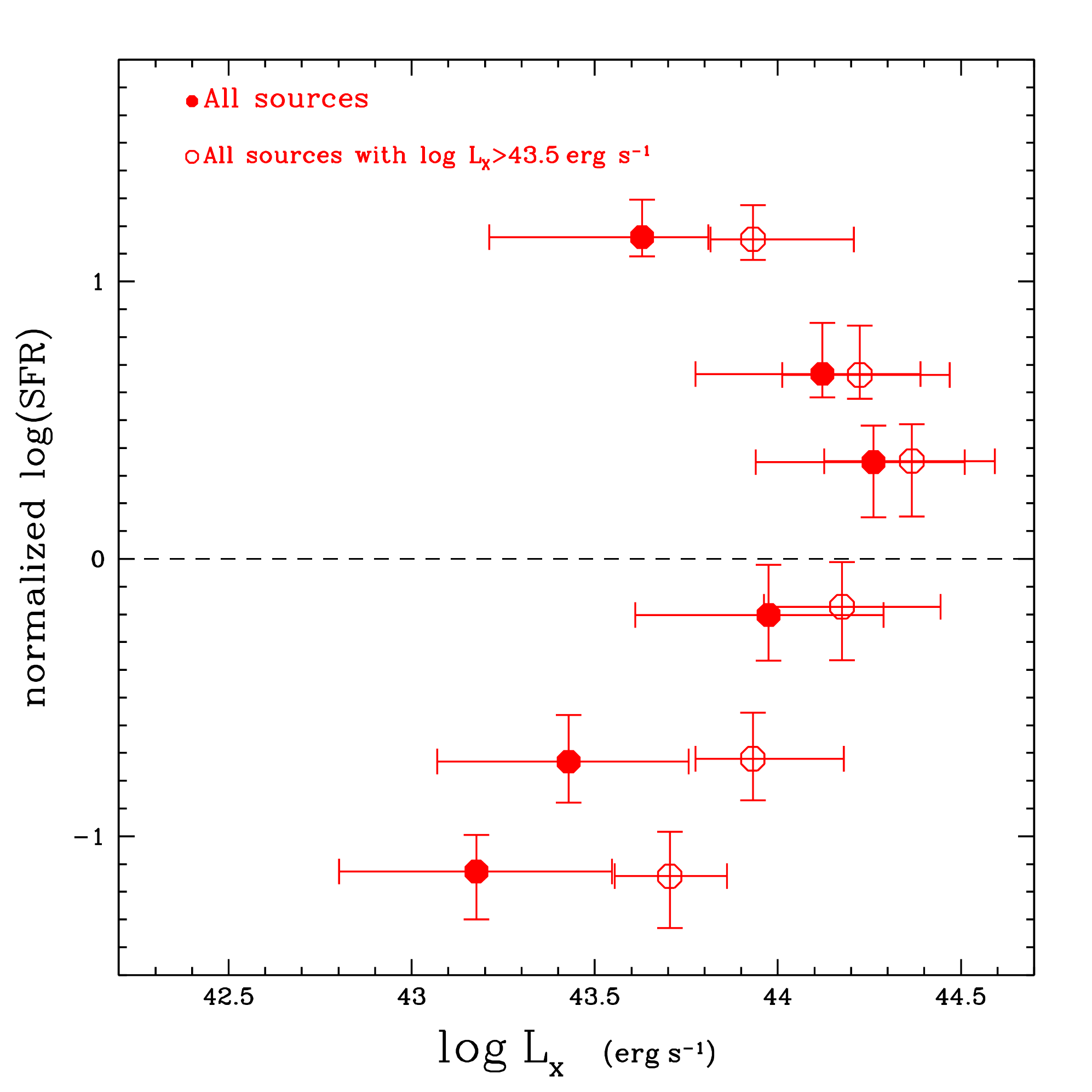}
  \label{fig:sub1}
\end{subfigure}%
\begin{subfigure}{.5\textwidth}
  \centering
  \includegraphics[width=1.03\linewidth]{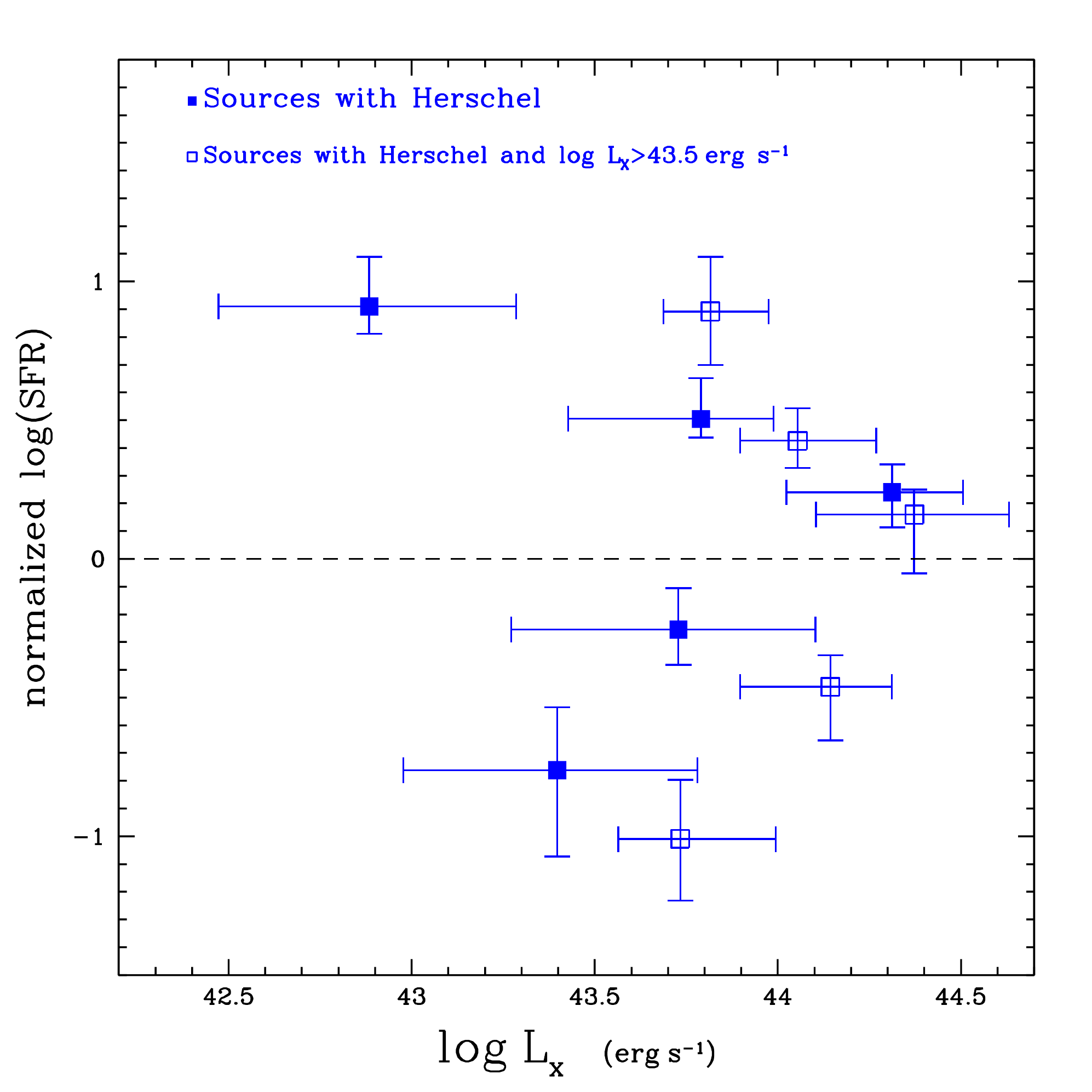}
  \label{fig:sub2}
\end{subfigure}
\caption{Binned results of the normalised SFR as a function of the median AGN X-ray luminosity, in bins of normalized SFR. {\bf{Left}}: Filled circles present the measurements for the 3336 sources. Open squares show our estimates when we restrict our sample to those X-ray AGN with  $\log \, L_X (\rm 2-10\,keV)  > 43.5$\, erg/s to account for the Lx-z incompleteness of our dataset (see Fig. \ref{fig_lx_z} and text for more details). {\bf{Right}}: Our results for the 608/3336 AGN with available Herschel photometry (filled squares). Open squares show our measurements for the 381 sources with Herschel photometry and  $\log \, L_X (\rm 2-10\,keV)  > 43.5$\, erg/s. In both panels the dashed line corresponds to AGN in the main sequence. Above this line, AGN have an enhanced SFR compared to star formation main-sequence galaxies of the same stellar mass and redshift. Below the dashed line, AGN have an SFR that is suppressed compared to main-sequence galaxies of the same mass and redshift. Based on our results, the AGN enhances the star formation of its host galaxy when the latter lies below the main sequence (below the dashed line) and quenches the star formation of the galaxy it lives in when the host lies above the main sequence (above the dashed line).}
\label{Main}
\end{figure*}

In this section, we use the X-ray luminosity as a proxy of the AGN power to study how an active super-massive black hole affects the star formation of its host galaxy. For this purpose we estimate the dependence of SFR on X-ray luminosity (Section 4.1). Then, we divide the SFR of the galaxy by its stellar mass to derive the specific SFR and study its dependence on the AGN power (Section 4.2). Previous works have estimated the evolution of SFR with M$_{\star}$ for star-forming galaxies \citep[e.g.][]{Schreiber2015}. We perform a similar analysis using our X-ray sources and compare our findings with those from galaxy studies (Section 4.3). To facilitate a better comparison with previous works \citep[e.g.][]{Lanzuisi2017, Brown2018}, we follow their analysis and do not account for the Lx-z incompleteness of our sample (see Fig. \ref{fig_lx_z}). However, we split our sample into redshift bins, which significantly reduced  the effect of incompleteness on our estimates. Finally, in Section 4.4 we distinguish the effect of stellar mass and redshift to test how this affects the dependence of SFR on L$_X$ found in Section 4.1.

\subsection{SFR vs. Lx}

In this section, we study how the X-ray luminosity, used as a proxy of the AGN power, affects the SFR of the host galaxy. SFRs are estimated by CIGALE through SED fitting. L$_X$ (observed) are estimated in the hard energy band ($2-10$\,keV) using the available flux estimates of the sources. Our measurements are presented in the left panel of Fig. \ref{SFR}. Individual sources are shown with small circles and squares for sources with and without Herschel photometry, respectively. We also compute median L$_X$ in bins of SFR, indicated by the filled squares (median SFR values are shown). Our results show a dependence of the SFR on L$_X$ in the whole redshift and luminosity range spanned by our sample. 

In the same figure, we also plot the binned SFR versus L$_X$ measurements from \cite{Brown2018} (polygons) and \cite{Lanzuisi2017} (triangles). \cite{Brown2018} used 703 AGN with $\rm log~L_X (\rm 2-10~keV)= (42-46)\,erg~s^{-1}$ at 0.1 < z < 5 from the Chandra XBo$\ddot{\rm{o}}$tes X-ray Survey and found a dependence of the star formation of the host galaxy on X-ray luminosity. Their SFR measurements are consistent with our estimates. \cite{Lanzuisi2017} used 692 AGN from the COSMOS field in the redshift range 0.1<z<4. Their results are in qualitative agreement with our findings, that is, they detected a dependence of the SFR on X-ray luminosity at all redshifts and luminosities. However, their SFR calculations appear higher than our results and those from \cite{Brown2018}.

Next, in the right panel of Fig. \ref{SFR} we compare our observational SFR versus L$_X$ results with the theoretical predictions of the model presented in \cite{Hickox2014}. In this model a population of star-forming galaxies across a range of redshifts from 0 to 2 is created, with a redshift-dependent distribution in SFR taken from the far-IR. Then, the far-IR luminosity is converted into SFR, and each galaxy is assigned an average BH accretion rate. Finally, the instantaneous accretion rate is converted into a bolometric AGN luminosity \citep[for more details, see][]{Hickox2014}. The curves are colour-coded based on the redshift bins. Although our results agree with the theoretical curves in the sense that the average estimates of the binned measurements are consistent with the model, our measurements show a stronger dependence of the SFR on Lx.

\subsection{sSFR vs. Lx}

In this section we explore the dependence of the sSFR, which is defined as the ratio of the SFR to the M$_*,$ on X-ray luminosity. In Fig. \ref{sSFRLx} we plot the results of our measurements. Squares indicate the average binned sSFR as a function of the mean L$_X$ of each bin for different redshift ranges. We chose to use mean sSFR and L$_X$ values instead of median to facilitate comparison with previous studies. Specifically, we overlaid the best-fit lines from the \cite{Rovilos2012}. \cite{Rovilos2012} used X-ray data from the 3Ms CDFS XMM-{\it{Newton}} survey and found a significant correlation between the sSFR and X-ray luminosity for redshifts $z>1$. They did not detect a significant correlation at lower redshifts, however. Our X-ray sample consists of more luminous sources than the Rovilos et al. dataset because of the large area of the XMM-XXL field used in our study. Towards this end, we extrapolated the best-fit lines of \cite{Rovilos2012} to higher X-ray luminosities (dashed lines). Our binned measurements agree well with their results at $1<\rm{z}<2.455$. At higher redshifts ($\rm{z}>2.5$), our estimates (cyan point) appear lower than the Rovilos et al. measurements. However, as previously described, their sample spans lower redshift and X-ray luminosities than ours, and only a small fraction of their sources resides at $z>2.5$. At lower redshifts, our sSFR measurements are lower but agree statistically with the Rovilos et al. sSFR estimates. Our measurements clearly show a dependence of the sSFR on redshift. In our lowest redshift bin, although the individual measurements have a large scatter, there is a hint of a mild sSFR dependence on L$_X$, even at $z<1$.

\subsection{Evolution of SFR with M$_\star$}

In this section, we examine the evolution of the average SFR of the host galaxies of X-ray AGN with stellar mass and redshift. The motivation is to compare this evolution for X-ray AGN and star-forming galaxies. \cite{Schreiber2015} used a sample of star-forming galaxies in four extragalactic fields, the GOODS North, GOODS South, UDS, and COSMOS obtained within the GOODS Herschel and CANDELS Herschel key programs. Their analysis revealed a universal, nearly linear slope of the $\log\,(SFR)-\log(M_\star)$ relation, with evidence for a flattening at high masses ($\log (M_{\star} / M_{\odot})>10.5$) that is less prominent at higher redshifts and almost vanishes at $z>2$. 

Fig. \ref{Mstellar} presents our measurements for the evolution of the average SFR of X-ray AGN with average stellar mass and redshift (filled squares). Average values were chosen to allow a fair comparison with the results of \cite{Schreiber2015} (open squares). Although our mean SFR estimates for the X-ray AGN are higher than the SFR of star-forming galaxies, with the exception of low redshifts (z$<0.7$), our results agree very well with the trends found by Schreiber et al. for the star-forming galaxies. Specifically, at low redshifts (z$<1.8$), SFR increases with M$_*$ for low stellar masses and then reaches a plateau for higher M$_\star$. At higher redshifts, the SFR increases nearly linearly with M$_\star$.

\subsection{Disentangling the effects of M$_\star$ and redshift on the SFR}
\label{sec_main}

Motivated by the results of the previous sections, that is, the strong dependence of SFR on M$_\star$ and redshift, we disentangle the effects of these parameters on SFR. Towards this end, we compare the SFR of our X-ray AGN with the SFR of main-sequence galaxies with the same stellar mass and redshift. The latter is estimated using Equation 9 in \cite{Schreiber2015}:

\begin{eqnarray}
\log_{10}({\rm SFR}_{\rm MS} [{\rm M}_\odot / {\rm yr}]) =  m - m_0 + a_0\,r \hspace{2.5cm} \nonumber \\
\hspace{0.9cm} - a_1 \, \big[{\rm max}(0, m - m_1 - a_2\,r)\big]^2\,,
\label{EQ:sfrms}
\end{eqnarray}
with $m_0 = 0.5 \pm 0.07$, $a_0 = 1.5 \pm 0.15$, $a_1 = 0.3 \pm 0.08$, $m_1 = 0.36 \pm 0.3,$ and $a_2 = 2.5 \pm 0.6$. r and m are defined as $r \equiv \log_{10}(1+z)$ and $m \equiv \log_{10}(M_\ast / 10^{9}\,{\rm M}_\odot).$

We then define the normalized SFR as the ratio of the SFR of X-ray AGN to the SFR of galaxies in the main sequence. Our measurements in bins of normalized SFR are presented in Fig. \ref{Main}. To test whether the incompleteness of our sample in the Lx-z plane (see Section \ref{sec_final_sample}) affects our measurements, we also applied a luminosity cut to our data: $\log \, L_X (\rm 2-10\,keV) > 43.5$\, erg/s. These measurements are presented with open symbols. For AGN below the galaxy main sequence, that is, below the dashed line, we note that as the AGN power (L$_X$) increases, the SFR$_{norm}$ increases. For AGN above the main sequence, as we move to more powerful AGN (higher X-ray luminosities), the SFR$_{norm}$ decreases. Based on our results, the picture that emerges is that AGN enhances the star formation of its host galaxy when the latter lies below the main sequence (below the dashed line in Fig. \ref{Main}) and quenches the star formation of the galaxy it lives in when the host lies above the main sequence (above the dashed line). Therefore, the effect of AGN on the SFR of the host galaxy depends on the location of the galaxy relative to the main sequence. This trend is still detectable when we restrict our sample to the more luminous sources, that is,  $\log \, L_X (\rm 2-10\,keV) > 43.5$\, erg/s, to account for the lack of faint sources at high redshifts (open symbols). Based on these tests, we conclude that the observed trend cannot be attributed to a selection bias that affects our measurements. Instead, it is the result of disentangle the effect of M$_\star$ and redshift from the SFR of galaxies that host X-ray AGN.

\section{Summary and conclusions}

Using data from the XMM-XXL North and X-ATLAS fields that cover more than 50\,deg$^2$, we have composed the largest  X-ray AGN sample in the literature to perform an SED fitting analysis and study the effect of X-ray AGN on the SFR of their host galaxies. Our sample consists of 3,336 X-ray sources, 608 of which have available Herschel photometry and 1,872/3,336 have spectroscopic redshifts. We estimated star-formation rates and stellar mass using the CIGALE code. Using far-IR photometry (Herschel) when available, we derived a relation and calibrated the SFR estimates of the sources without Herschel observations. 

Our analysis reveals a dependence of the SFR of AGN host galaxies on the AGN power (X-ray luminosity) at all redshift and luminosity ranges spanned by our samples, that is, $0.3<z<3$ and $41< \log \, L_X (\rm 2-10\,keV) < 45.5$\, erg/s (Fig. \ref{SFR}). This result agrees with recent observation studies that used large X-ray samples, that is, \cite{Brown2018} and \cite{Lanzuisi2017}. However, the Lanzuisi et al. measurements appear higher than the result of our study and that of Brown et al. The comparison of our findings with theoretical predictions \citep{Hickox2014} shows that although our measurements show a stronger dependence of the SFR on L$_X$, our results are consistent with these predictions. 

Furthermore, we find that the sSFR of the AGN host galaxies increases with redshift, while there are indications that the X-ray luminosity enhances the sSFR of the host galaxy. Our results agree with the findings from \cite{Rovilos2012}. We also find hints that the sSFR-L$_x$ dependence holds at low redshifts (z$<1$; Fig. \ref{sSFRLx}).

Following the work that has been done on star-forming galaxies \citep{Schreiber2015}, we studied the evolution of SFR with M$_\star$ and redshift for our X-ray sources. Our results show that the SFR-M$_\star$ evolution for the X-ray AGN shows similar trends as those found for star-forming galaxies (Fig. \ref{Mstellar}). Specifically, at low redshifts (z $<1.8$), the SFR increases with M$_\star$ for low stellar masses and then reaches a plateau for higher M$_\star$. At higher redshifts, the SFR increases nearly linearly with stellar mass. We also find that the mean SFRs of AGN are higher than the SFR of star-forming galaxies at z $>0.7$.

Prompted by the SFR-M$_\star$ evolution with redshift, we disentangled the effects of M$_\star$ and redshift on the SFR. For this purpose, we used the formula of \cite{Schreiber2015} to estimate the SFR of main-sequence galaxies that have the same stellar mass and redshift with our X-ray AGN. We reduced the effects of incompleteness of our sample on our measurements by excluding low X-ray luminosity sources from our AGN dataset. Our analysis reveals that the AGN enhances the star formation of its host galaxy when the latter lies below the main sequence and quenches the star formation of the galaxy it lives in when the host lies above the main sequence. Therefore, the effect of AGN on the SFR of the host galaxy depends on the location of the galaxy relative to the main sequence (Fig. \ref{Main}). 

Our study shows that discrepancies found among the results of previous studies regarding the effect of AGN on the host galaxy star-formation can be mitigated when large X-ray samples are used in the analysis. Even more importantly, it is essential to distinguish the effects of stellar mass and redshift on SFR when we study the role of the AGN on the star formation of its host galaxy. When this is taken into account, the complex relation between the AGN and the host galaxy is revealed. AGN can either enhance or quench the star formation, depending on how powerful the AGN is. 

\begin{acknowledgements}
The authors are grateful to the anonymous referee for helpful comments. GM acknowledges support of this work by the PROTEAS II project (MIS 5002515), which is implemented under the “Reinforcement of the Research and Innovation Infrastructure” action, funded by the ”Competitiveness, Entrepreneurship and Innovation” operational programme (NSRF 2014-2020) and co-financed by Greece and the European Union (European Regional Development Fund). ACR acknowledges financial support by the European Space Agency (ESA) under the PRODEX program.
\\
This research has made use of data obtained from the 3XMM XMM-\textit{Newton} 
serendipitous source catalogue compiled by the 10 institutes of the XMM-\textit{Newton} 
Survey Science Centre selected by ESA.
\\
This work is based on observations made with XMM-\textit{Newton}, an ESA science 
mission with instruments and contributions directly funded by ESA Member States 
and NASA. 
\\
Funding for the Sloan Digital Sky Survey IV has been provided by the Alfred P. Sloan Foundation, the U.S. Department of Energy Office of Science, and the Participating Institutions. SDSS-IV acknowledges
support and resources from the Center for High-Performance Computing at
the University of Utah. The SDSS web site is \url{www.sdss.org}.
\\
SDSS-IV is managed by the Astrophysical Research Consortium for the 
Participating Institutions of the SDSS Collaboration including the 
Brazilian Participation Group, the Carnegie Institution for Science, 
Carnegie Mellon University, the Chilean Participation Group, the French Participation Group, Harvard-Smithsonian Center for Astrophysics, 
Instituto de Astrof\'isica de Canarias, The Johns Hopkins University, 
Kavli Institute for the Physics and Mathematics of the Universe (IPMU) / 
University of Tokyo, Lawrence Berkeley National Laboratory, 
Leibniz Institut f\"ur Astrophysik Potsdam (AIP),  
Max-Planck-Institut f\"ur Astronomie (MPIA Heidelberg), 
Max-Planck-Institut f\"ur Astrophysik (MPA Garching), 
Max-Planck-Institut f\"ur Extraterrestrische Physik (MPE), 
National Astronomical Observatories of China, New Mexico State University, 
New York University, University of Notre Dame, 
Observat\'ario Nacional / MCTI, The Ohio State University, 
Pennsylvania State University, Shanghai Astronomical Observatory, 
United Kingdom Participation Group,
Universidad Nacional Aut\'onoma de M\'exico, University of Arizona, 
University of Colorado Boulder, University of Oxford, University of Portsmouth, 
University of Utah, University of Virginia, University of Washington, University of Wisconsin, 
Vanderbilt University, and Yale University.
\\
This publication makes use of data products from the Wide-field Infrared Survey 
Explorer, which is a joint project of the University of California, Los Angeles, 
and the Jet Propulsion Laboratory/California Institute of Technology, funded by 
the National Aeronautics and Space Administration. 
\\
The VISTA Data Flow System pipeline processing and science archive are described 
in \cite{Irwin2004}, \cite{Hambly2008} and \cite{Cross2012}. Based on 
observations obtained as part of the VISTA Hemisphere Survey, ESO Program, 
179.A-2010 (PI: McMahon). We have used data from the 3rd data release.
\\
\end{acknowledgements}

\bibliography{mybib}{}

\begin{thebibliography}{57}
\expandafter\ifx\csname natexlab\endcsname\relax\def\natexlab#1{#1}\fi

\bibitem[{Albareti {et~al.}(2015)Albareti, Comparat, Guti{\'e}rrez, Prada,
  P{\^a}ris, Schlegel, L{\'o}pez-Corredoira, Schneider, Manchado,
  Garc{\'\i}a-Hern{\'a}ndez, Petitjean, \& Ge}]{Albareti2015}
Albareti, F.~D., Comparat, J., Guti{\'e}rrez, C.~M., {et~al.} 2015, MNRAS, 452,
  4153

\bibitem[{{Alexander} \& {Hickox}(2012)}]{Alexander2012}
{Alexander}, D.~M. \& {Hickox}, R.~C. 2012, NewAR, 56, 93

\bibitem[{Baldry {et~al.}(2010)Baldry, Robotham, Hill, Driver, Liske, Norberg,
  Bamford, Hopkins, Loveday, Peacock, Cameron, Croom, Cross, Doyle, Dye, Frenk,
  Jones, van Kampen, Kelvin, Nichol, Parkinson, Popescu, Prescott, Sharp,
  Sutherland, Thomas, \& Tuffs}]{Baldry2010}
Baldry, I.~K., Robotham, A. S.~G., Hill, D.~T., {et~al.} 2010, MNRAS, 404, 86

\bibitem[{Birnboim \& Dekel(2003)}]{Birnboim2003}
Birnboim, Y. \& Dekel, A. 2003, MNRAS, 345, 349

\bibitem[{Bonfield {et~al.}(2011)}]{Bonfield2011}
Bonfield, D.~G. {et~al.} 2011, MNRAS, 416, 13

\bibitem[{Boyle \& Terlevich(1998)}]{Boyle1998}
Boyle, B.~J. \& Terlevich, R.~J. 1998, MNRAS, 293, 49

\bibitem[{Brown {et~al.}(2018)Brown, Nayyeri, Cooray, Ma, {Hickox}, \&
  Azadi}]{Brown2018}
Brown, A., Nayyeri, H., Cooray, A., {et~al.} 2018, eprint arXiv:1801.02233

\bibitem[{Bruzual \& Charlot(2003)}]{Bruzual_Charlot2003}
Bruzual, G. \& Charlot, S. 2003, MNRAS, 344, 1000

\bibitem[{{Calzetti} {et~al.}(2000){Calzetti}, {Armus}, {Bohlin}, {Kinney},
  {Koornneef}, \& {Storchi-Bergmann}}]{Calzetti2000}
{Calzetti}, D., {Armus}, L., {Bohlin}, R.~C., {et~al.} 2000, ApJ, 533, 682

\bibitem[{Carrasco~Kind \& {Brunner}(2013)}]{Kind2013}
Carrasco~Kind, M. \& {Brunner}, R.~J. 2013, MNRAS, 432, 1483

\bibitem[{{Ciesla} {et~al.}(2015)}]{Ciesla2015}
{Ciesla}, L. {et~al.} 2015, A\&A, 576, 19

\bibitem[{Cross {et~al.}(2012)}]{Cross2012}
Cross, N. J.~G. {et~al.} 2012, A\&A, 548, 21

\bibitem[{{Dale} {et~al.}(2014){Dale}, {Helou}, {Magdis}, {Armus},
  {D{\'{\i}}az-Santos}, \& {Shi}}]{Dale2014}
{Dale}, D.~A., {Helou}, G., {Magdis}, G.~E., {et~al.} 2014, ApJ, 784, 83

\bibitem[{Dalton {et~al.}(2006)Dalton, Caldwell, Ward, Whalley, Woodhouse,
  Edeson, Clark, Beard, Gallie, Todd, Strachan, Bezawada, Sutherland, \&
  Emerson}]{Dalton2006}
Dalton, G.~B., Caldwell, M., Ward, A.~K., {et~al.} 2006, SPIE, 6269

\bibitem[{{Di Matteo} {et~al.}(2008){Di Matteo}, {Colberg}, {Springel},
  {Hernquist}, \& {Sijacki}}]{DiMatteo2008}
{Di Matteo}, T., {Colberg}, J., {Springel}, V., {Hernquist}, L., \& {Sijacki},
  D. 2008, ApJ, 676, 33

\bibitem[{{Di Matteo} {et~al.}(2005){Di Matteo}, {Springel}, \&
  {Hernquist}}]{DiMatteo2005}
{Di Matteo}, T., {Springel}, V., \& {Hernquist}, L. 2005, Nature, 433, 604

\bibitem[{Driver {et~al.}(2011)Driver, Hill, Kelvin, Robotham, Liske, Norberg,
  Baldry, Bamford, Hopkins, Loveday, Peacock, Andrae, Bland-Hawthorn, Brough,
  Brown, Cameron, Ching, Colless, Conselice, Croom, Cross, de~Propris, Dye,
  Drinkwater, Ellis, Graham, Grootes, Gunawardhana, Jones, van Kampen,
  Maraston, Nichol, Parkinson, Phillipps, Pimbblet, Popescu, Prescott,
  Roseboom, Sadler, Sansom, Sharp, Smith, Taylor, Thomas, Tuffs, Wijesinghe,
  Dunne, Frenk, Jarvis, Madore, Meyer, Seibert, Staveley-Smith, Sutherland, \&
  Warren}]{Driver2011}
Driver, S.~P., Hill, D.~T., Kelvin, L.~S., {et~al.} 2011, MNRAS, 413, 971

\bibitem[{Duras {et~al.}(2017)}]{Duras2017}
Duras, F. {et~al.} 2017, A\&A, 604, 19

\bibitem[{Eales {et~al.}(2010)Eales, Dunne, Clements, Cooray, De~Zotti, Dye,
  Ivison, Jarvis, Lagache, Maddox, Negrello, Serjeant, Thompson, Van~Kampen,
  Amblard, Andreani, Baes, Beelen, Bendo, Benford, Bertoldi, Bock, Bonfield,
  Boselli, Bridge, Buat, Burgarella, Carlberg, Cava, Chanial, Charlot,
  Christopher, Coles, Cortese, Dariush, da~Cunha, Dalton, Danese, Dannerbauer,
  Driver, Dunlop, Fan, Farrah, Frayer, Frenk, Geach, Gardner, Gomez,
  Gonz{\'a}lez-Nuevo, Gonz{\'a}lez-Solares, Griffin, Hardcastle,
  Hatziminaoglou, Herranz, Hughes, Ibar, Jeong, Lacey, Lapi, Lawrence, Lee,
  Leeuw, Liske, L{\'o}pez-Caniego, M{\"u}ller, Nandra, Panuzzo, Papageorgiou,
  Patanchon, Peacock, Pearson, Phillipps, Pohlen, Popescu, Rawlings, Rigby,
  Rigopoulou, Robotham, Rodighiero, Sansom, Schulz, Scott, Smith, Sibthorpe,
  Smail, Stevens, Sutherland, Takeuchi, Tedds, Temi, Tuffs, Trichas, Vaccari,
  Valtchanov, van~der Werf, Verma, Vieria, Vlahakis, \& White}]{Eales2010}
Eales, S., Dunne, L., Clements, D., {et~al.} 2010, PASP, 122, 499

\bibitem[{Emerson {et~al.}(2006)Emerson, McPherson, \&
  Sutherland}]{Emerson2006}
Emerson, J., McPherson, A., \& Sutherland, W. 2006, Msngr, 126, 41

\bibitem[{{Ferrarese} \& {Merritt}(2000)}]{Ferrarese2000}
{Ferrarese}, L. \& {Merritt}, D. 2000, ApJ, 539, 9

\bibitem[{{Fritz} {et~al.}(2006){Fritz}, {Franceschini}, \&
  {Hatziminaoglou}}]{Fritz2006}
{Fritz}, J., {Franceschini}, A., \& {Hatziminaoglou}, E. 2006, MNRAS, 166, 767

\bibitem[{Gabor {et~al.}(2010)Gabor, Dav{\'e}, Finlator, \&
  Oppenheimer}]{Gabor2010}
Gabor, J.~M., Dav{\'e}, R., Finlator, K., \& Oppenheimer, B.~D. 2010, MNRAS,
  407, 749

\bibitem[{{Georgakakis} {et~al.}(2008){Georgakakis}, {Gerke}, {Nandra},
  {Laird}, {Coil}, {Cooper}, \& {Newman}}]{Georgakakis2008}
{Georgakakis}, A., {Gerke}, B.~F., {Nandra}, K., {et~al.} 2008, MNRAS, 391, 183

\bibitem[{{Georgakakis} \& {Nandra}(2011)}]{Georgakakis_Nandra2011}
{Georgakakis}, A. \& {Nandra}, K. 2011, ArXiv: 1101.4943
  [\eprint[arXiv]{1101.4943}]

\bibitem[{{Georgakakis} {et~al.}(2017)}]{Georgakakis2017}
{Georgakakis}, A. {et~al.} 2017, MNRAS, 469, 3232

\bibitem[{Hambly {et~al.}(2008)}]{Hambly2008}
Hambly, N.~C. {et~al.} 2008, MNRAS, 384, 637

\bibitem[{{Harrison} {et~al.}(2012)}]{Harrison2012}
{Harrison}, C.~M. {et~al.} 2012, ApJL, 760, 5

\bibitem[{Hickox {et~al.}(2014)Hickox, Mullaney, Alexander, Chen, Civano, \&
  Goulding}]{Hickox2014}
Hickox, R.~C., Mullaney, J.~R., Alexander, D.~M., {et~al.} 2014, ApJ, 782, 11

\bibitem[{{Hopkins} {et~al.}(2008){Hopkins}, {Hernquist}, {Cox}, \&
  {Keres}}]{Hopkins2008a}
{Hopkins}, P.~F., {Hernquist}, L., {Cox}, T.~J., \& {Keres}, D. 2008, ApJS,
  175, 356

\bibitem[{{Hopkins} {et~al.}(2006){Hopkins}, {Hernquist}, {Cox}, {Robertson},
  {Di Matteo}, \& {Springel}}]{Hopkins2006a}
{Hopkins}, P.~F., {Hernquist}, L., {Cox}, T.~J., {et~al.} 2006, ApJ, 639, 700

\bibitem[{{Hopkins} {et~al.}(2010)}]{Hopkins2010}
{Hopkins}, P.~F. {et~al.} 2010, ApJ, 724, 915

\bibitem[{Irwin {et~al.}(2004)}]{Irwin2004}
Irwin, M.~J. {et~al.} 2004, SPIE, 5493, 411

\bibitem[{Kere{\v s} {et~al.}(2009)Kere{\v s}, Katz, Dav{\'e}, Fardal, \&
  Weinberg}]{Keres2009}
Kere{\v s}, D., Katz, N., Dav{\'e}, R., Fardal, M., \& Weinberg, D.~H. 2009,
  MNRAS, 396, 2332

\bibitem[{{Lanzuisi} {et~al.}(2017)}]{Lanzuisi2017}
{Lanzuisi}, G. {et~al.} 2017, A\&A, 602, 13

\bibitem[{Liu {et~al.}(2016)Liu, Merloni, Georgakakis, Menzel, Buchner, Nandra,
  Salvato, Shen, Brusa, \& Streblyanska}]{Liu2016}
Liu, Z., Merloni, A., Georgakakis, A., {et~al.} 2016, MNRAS, 459, 1602

\bibitem[{{Lusso} {et~al.}(2012)}]{Lusso2012}
{Lusso}, E. {et~al.} 2012, MNRAS, 425, 623

\bibitem[{{Lutz} {et~al.}(2010)}]{Lutz2010}
{Lutz}, D. {et~al.} 2010, ApJ, 712, 1287

\bibitem[{Magorrian {et~al.}(1998)}]{Magorrian2008}
Magorrian, J. {et~al.} 1998, AJ, 115, 2285

\bibitem[{Menzel {et~al.}(2016)Menzel, Merloni, Georgakakis, Salvato, Aubourg,
  Brandt, Brusa, Buchner, Dwelly, Nandra, P{\^a}ris, Petitjean, \&
  Schwope}]{Menzel2016}
Menzel, M.-L., Merloni, A., Georgakakis, A., {et~al.} 2016, MNRAS, 457, 110

\bibitem[{{Mountrichas} {et~al.}(2017){Mountrichas}, {Corral}, {Masoura},
  {Georgantopoulos}, {Ruiz}, {Georgakakis}, {Carrera}, \&
  {Fotopoulou}}]{Mountrichas2017b}
{Mountrichas}, G., {Corral}, A., {Masoura}, V.~A., {et~al.} 2017, A\&A, 608, 39

\bibitem[{{Noll} {et~al.}(2009)}]{Noll2009}
{Noll}, S. {et~al.} 2009, A\&A, 507, 1793

\bibitem[{{Oliver} {et~al.}(2012)}]{Oliver2012}
{Oliver}, S.~J. {et~al.} 2012, MNRAS, 424, 1614

\bibitem[{{Page} {et~al.}(2012)}]{Page2012}
{Page}, M.~J. {et~al.} 2012, nat, 485, 213

\bibitem[{{Pierre} {et~al.}(2016)}]{Pierre2016}
{Pierre}, M. {et~al.} 2016, A\&A, 592, 1

\bibitem[{Pineau {et~al.}(2017)}]{Pineau2017}
Pineau, F.~X. {et~al.} 2017, A\&A, 597, 28

\bibitem[{Ranalli {et~al.}(2015)Ranalli, Georgantopoulos, Corral, Koutoulidis,
  Rovilos, Carrera, Akylas, Del~Moro, Georgakakis, Gilli, \&
  Vignali}]{Ranalli2015}
Ranalli, P., Georgantopoulos, I., Corral, A., {et~al.} 2015, A\&A, 577, 10

\bibitem[{Rigby {et~al.}(2011)Rigby, Maddox, Dunne, Negrello, Smith,
  Gonz{\'a}lez-Nuevo, Herranz, L{\'o}pez-Caniego, Auld, Buttiglione, Baes,
  Cava, Cooray, Clements, Dariush, de~Zotti, Dye, Eales, Frayer, Fritz,
  Hopwood, Ibar, Ivison, Jarvis, Panuzzo, Pascale, Pohlen, Rodighiero,
  Serjeant, Temi, \& Thompson}]{Rigby2011}
Rigby, E.~E., Maddox, S.~J., Dunne, L., {et~al.} 2011, MNRAS, 415, 2336

\bibitem[{{Rovilos} {et~al.}(2012)}]{Rovilos2012}
{Rovilos}, E. {et~al.} 2012, A\&A, 546, 16

\bibitem[{{Salvato} {et~al.}(2011)}]{Salvato2011}
{Salvato}, M. {et~al.} 2011, ApJ, 742, 61

\bibitem[{Schreiber {et~al.}(2015)}]{Schreiber2015}
Schreiber, C. {et~al.} 2015, A\&A, 575, 29

\bibitem[{Sijacki {et~al.}(2007)Sijacki, Springel, \& Di~Matteo}]{Sijacki2007}
Sijacki, D., Springel, V., \& Di~Matteo, T. 2007, MNRAS, 380, 877

\bibitem[{{Springel} {et~al.}(2005)}]{Springel2005}
{Springel}, V. {et~al.} 2005, nat, 435, 629

\bibitem[{{Sutherland}(1992)}]{Sutherland_and_Saunders1992}
{Sutherland}, W. \&~{Saunders}, W. 1992, MNRAS, 259, 413

\bibitem[{{Symeonidis} {et~al.}(2016){Symeonidis}, {Giblin}, {Page}, {Pearson},
  {Bendo}, {Seymour}, \& {Oliver}}]{Symeonidis2016}
{Symeonidis}, M., {Giblin}, B.~M., {Page}, M.~J., {et~al.} 2016, MNRAS, 459,
  257

\bibitem[{Valiante {et~al.}(2016)Valiante, Smith, Eales, Maddox, Ibar, Hopwood,
  Dunne, Cigan, Dye, Pascale, Rigby, Bourne, Furlanetto, \&
  Ivison}]{Valiante2016}
Valiante, E., Smith, M. W.~L., Eales, S., {et~al.} 2016, MNRAS, 462, 3146

\bibitem[{{Wright} {et~al.}(2010){Wright}, {Eisenhardt}, {Mainzer}, {Ressler},
  {Cutri}, {Jarrett}, {Kirkpatrick}, {Padgett}, {McMillan}, {Skrutskie},
  {Stanford}, {Cohen}, {Walker}, {Mather}, {Leisawitz}, {Gautier}, {McLean},
  {Benford}, {Lonsdale}, {Blain}, {Mendez}, {Irace}, {Duval}, {Liu}, {Royer},
  {Heinrichsen}, {Howard}, {Shannon}, {Kendall}, {Walsh}, {Larsen}, {Cardon},
  {Schick}, {Schwalm}, {Abid}, {Fabinsky}, {Naes}, \& {Tsai}}]{Wright2010}
{Wright}, E.~L., {Eisenhardt}, P.~R.~M., {Mainzer}, A.~K., {et~al.} 2010, AJ,
  140, 1868

\end{thebibliography}
\bibliographystyle{aa}
\end{document}